\documentclass[iop,apjl,tighten]{emulateapj}
\usepackage{amssymb}
\usepackage{times}
\usepackage{epsfig}
\usepackage{graphics}
\usepackage{natbib}

\begin{document}

\title{A synchrotron self-Compton -- disk reprocessing model for optical/X-ray correlation in black hole X-ray binaries}

\shorttitle{Optical/X-ray correlation in black hole X-ray binaries}
\shortauthors{Veledina, Poutanen, \& Vurm}

\author{Alexandra Veledina,\altaffilmark{1} Juri Poutanen,\altaffilmark{1} and Indrek Vurm\altaffilmark{2}} 

\affil{$^1$Astronomy Division, Department of Physics, P.O. Box 3000, 
90014 University of Oulu, Finland; alexandra.veledina@oulu.fi, juri.poutanen@oulu.fi \\
$^2$Racah Institute of Physics, Hebrew University of Jerusalem, 91904 Jerusalem, Israel; indrek@phys.huji.ac.il}


\begin{abstract} 
\noindent 
Physical picture of the emission mechanisms operating in the X-ray binaries was put under question by the simultaneous
optical/X-ray observations with high time resolution.
The light curves of the two energy bands appeared to be connected and the cross-correlation functions observed in three black
hole binaries exhibited a complicated shape.
They show a dip of the optical emission a few seconds before the X-ray peak and the optical flare just after the X-ray peak.
This behavior could not be explained in terms of standard optical emission candidates 
(e.g., emission from the cold accretion disk  or a jet).
We propose a novel model, which explains the broadband optical to the X-ray spectra and the variability properties.
We suggest that the optical emission consists of two components: synchrotron radiation from the non-thermal electrons in the 
hot accretion flow and the emission produced by reprocessing of the X-rays in the outer part of the accretion disk. 
The first component is anti-correlated with the X-rays, while the second one is correlated, but delayed  and smeared relative to
the X-rays. 
The interplay of the components explains the complex shape of the cross-correlation function, the features 
in the optical power spectral density  as well as the time lags. 
\end{abstract}


\section{Introduction}

The physical processes giving rise to the broadband radio to X-ray spectra of accreting black holes are still under debate. 
There is a general consensus that the radio emission is associated with the jet and the X-rays are produced in the vicinity of 
the compact object.
However, the origin of the IR/optical/UV emission from the black holes in  low-mass X-ray binaries (LMXBs) 
is less certain.
The contribution from the companion star in such systems is usually too faint, and the optical spectra are most likely
connected to the accretion process onto the compact object and the X-ray radiation.
To understand the origin of the optical emission, timing analyses of simultaneous optical and X-ray observations were performed.
Such observations were for the first time carried out by \citet{Motch83} for the black-hole binary (BHB) GX 339--4.
Although the duration of the observations was too short for any confident conclusion, the computed optical/X-ray
cross-correlation function (CCF) revealed a complicated structure with a dip in the optical light curve, preceding the X-ray
peak (the so-called precognition dip), together with an optical peak lagging the X-rays.
Recently, CCFs were obtained from the much longer duration simultaneous observations in three LMXBs: XTE J1118+480
\citep{Kanbach01,HHC03}, Swift J1753.5--0127 \citep{DGS08,DSG11,HBM09} and GX 339--4 \citep{GMD08}, all manifesting similar
structure.
The observed behavior cannot be explained by a simple model with the optical radiation being produced by the
reprocessed X-ray emission \citep{HHC03}. 
This hypothesis also contradicts the fact that the observed autocorrelation function (ACF) of the optical radiation is narrower
than that of the X-rays \citep{Kanbach01,HHC03,DGS09,GDD10}.
On the other hand, study of the dependence of the CCF on the timescale of fluctuations in the light curve revealed that its
overall shape remains similar, but rescaled \citep{MBSK03,GDD10}, suggesting that the emission in the two energy bands is
intrinsically connected.

A number of possible mechanisms, producing such structure of the CCF, were proposed (see the discussion in
\citealt{DSG11} and references therein).
So far, the detailed calculations were made only in the model of \citet{MMF04}, where both X-ray and optical
emission are powered by the same magnetic energy reservoir.
The model qualitatively describes the CCF and ACF of XTE J1118+480, however, with the new data on Swift J1753.5--0127, the
applicability of the model was put under question.
It is also doubtful, that such an energetic magnetized zone is consistent with the accretion theories.
 
In this Letter, we suggest that the optical emission is a composition of two components.
The first one is coming from the synchrotron-emitting particles in the inner hot flow and is anti-correlated with the X-ray 
emission, which is produced by Comptonization of the synchrotron radiation.
The second one is originating from reprocessing of the X-rays by the outer parts of the cold accretion disk
\citep{DLHC99,GDP09,CDSG10}.
This component is (positively) correlated and delayed relative to the X-rays.
The presence of two components explains both the precognition dip and the delayed peak of the CCF.
The model is capable of explaining the entire optical to the X-ray spectrum of BHBs as well as their timing properties.

\begin{figure}
\plotone{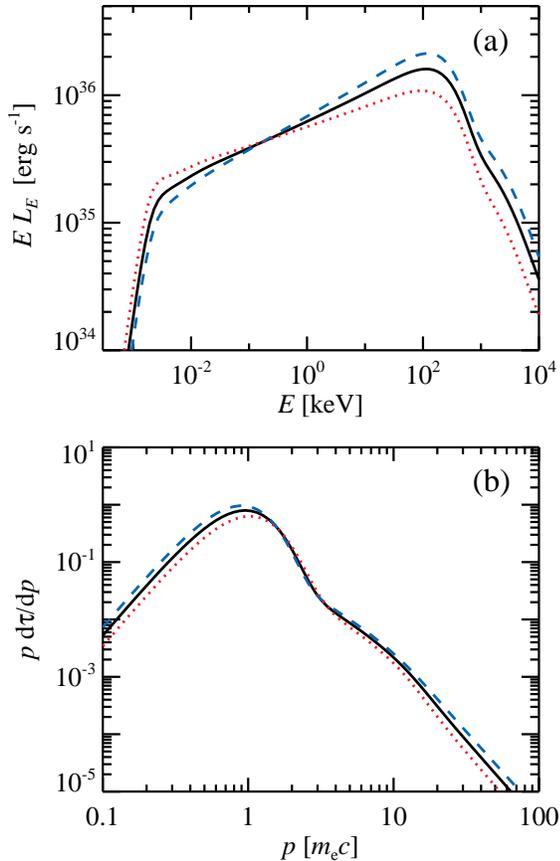}
\caption{
Simulated  (a)  photon spectra and  (b) electron distributions for the SSC model for hybrid plasmas.
Solid lines correspond to the fiducial parameter set (see the text), an increase of the accretion rate by  20\% hardens the
spectrum (blue dashed line) and a decrease of the accretion rate by  20\% softens the spectrum (red dotted line).
These spectra are associated with the inner hot flow and an additional contribution of the irradiated accretion disk is not
shown.
}\label{pic:hybComp}
\end{figure}

\section{Spectral model}\label{sect:specmod}

The X-ray spectra of BHBs in their hard state are well described by Comptonization on thermal electrons 
\citep[e.g.,][]{P98,ZG04}, however power-law tails detected in a number of sources \citep{ZGP01,McConnell02}
suggest the presence of non-thermal particle population in addition to the mostly thermal distribution \citep{PC98}.
These electrons contribute both to the high-energy part of the spectrum via Compton scattering and to the low-energy part by
synchrotron radiation.
Even a tiny fraction of non-thermal electrons dramatically increases the synchrotron luminosity \citep{WZ01} and thus makes it
a plausible source of seed photons for Comptonization.
The X-ray spectra of the hard state BHBs (and supermassive black holes) can be modeled in terms of synchrotron
self-Compton (SSC) mechanism in hybrid (thermal plus non-thermal) plasmas \citep{VP08,PV09,MB09,VVP11}.
The high-energy part of the spectrum is dominated by the Comptonized photons, and the low-energy part is determined by the
synchrotron emission, which can extend down to the IR/optical energy bands depending on the size of the emission region.

For simplicity, we consider a spherical emission region homogeneously filled with photons and electrons.
The region corresponds to a hot flow in the vicinity of the black hole \citep[see reviews in][]{NMQ98,Yuan07}, 
with the size limited by the inner radius of the (truncated) accretion disk. 
We consider the energy input in the form of injected electrons with a power-law spectrum 
$dN_{\rm e}/(dt\ d\gamma) \propto \gamma^{-\Gamma_{\rm inj}}$ extending between the Lorentz factors
$\gamma_{\min}$ and $\gamma_{\max}$.
The injection might result from magnetic reconnection or shock acceleration.
The main mechanisms responsible for particle cooling and thermalization and formation of the spectra are cyclo-synchrotron
emission and absorption, Compton scattering and electron-electron Coulomb collisions.
To find self-consistent photon spectra and  electron distributions, we solve a set of 
kinetic equations for photons, electrons, and positrons using the numerical code developed by \citet{VP09}.

A good description of the broadband spectra of hard-state BHBs can be achieved with the following parameters
\citep[see][]{PV09}: the Thomson optical depth $\tau$=1.0, the magnetic field $B=3\times10^5$~G and the total
luminosity of the region $L=10^{37}$~erg~s$^{-1}$.
The parameters of the injection function are $\Gamma_{\rm inj}=3$, $\gamma_{\min} = 1.0$, and $\gamma_{\max} = 10^3$.
The size of the emission region $R$ is assumed to be 30 Schwarzschild radii, which corresponds to $9\times10^7$~cm for a 
$10 M_{\odot}$ black hole.
For this set of parameters, the synchrotron self-absorption frequency falls in the optical energy band (1$-$10 eV). 
The resulting spectrum and the electron distribution are shown in Figure~\ref{pic:hybComp} (solid line).

The broadband spectra are expected to vary due to the changes of the mass accretion rate $\dot m$. 
We assume that the luminosity varies as $L \propto \dot m$, the optical depth as $\tau \propto \dot m$, and the size of
the hot flow (inner cold disk radius) as $R \propto \dot m^{-4/3}$ \citep{RC00}.
We assume that the magnetic field and the injection index $\Gamma_{\rm inj}$ are constant.
With an increase of the accretion rate, the X-ray luminosity increases, whereas the optical radiation drops due to the increased
synchrotron self-absorption and vice versa (dashed and dotted lines in Figure~\ref{pic:hybComp}).
The resulting X-ray spectral behavior with hardening during the X-ray flares and softening during the X-ray dips is consistent
with the observations \citep{GDD10}.
The decrease of the synchrotron radiation with an increase of the mass accretion rate was also found to take place in 
the advective disk models \citep{YCN05}.
Thus, the optical synchrotron luminosity is expected to anti-correlate with the X-rays.

\begin{figure*}
\plotone{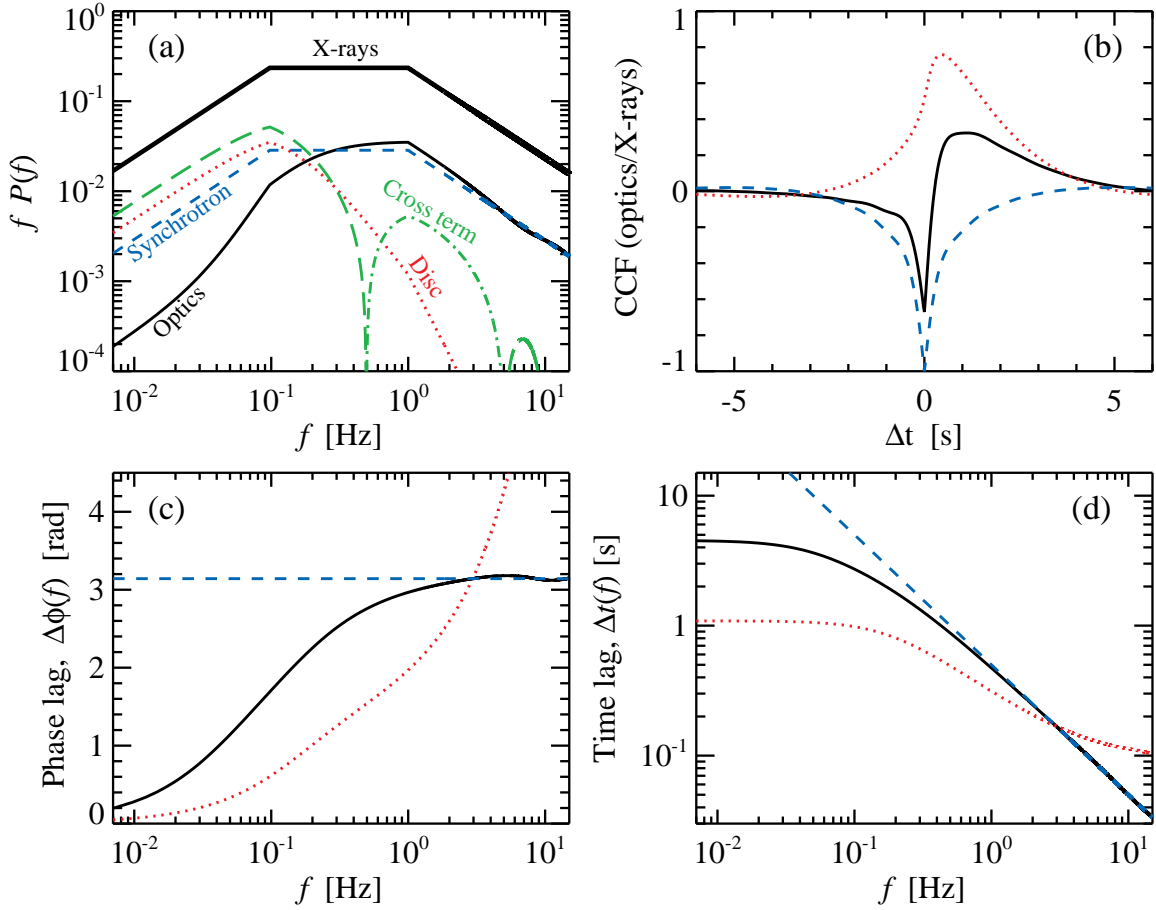}
\caption{
Results of simulations for the model with the exponential disk response function with $\tau_1=0.1$ and $\tau_2=1.0$, and
$r_{\rm ds} = 1.3$.
(a) The PSDs of the X-ray  (double-broken PSD, thick solid line) and optical light curves (thin solid line). 
Three terms contributing to the optical PSD (see Equation (\ref{eq:psdo})) are also shown: synchrotron (dashed line), disk
(dotted line), and the cross term (positive contribution is shown by the dot-dashed line and the negative contribution by
the long-dashed line).
(b) The CCFs, (c) the phase lags, and (d) the time lags corresponding to the synchrotron/X-rays ($r_{\rm ds} = 0$, dashed
line), reprocessed disk component/X-rays ($r_{\rm ds} \rightarrow \infty$, dotted line), and the combined optical/X-rays (solid
line). 
}\label{pic:result}
\end{figure*}

\section{Timing model}
\label{sect:timemod}

\subsection{Formalism}
\label{sect:time_method}
 
The X-ray and the optical light curves can be represented as a sum of a constant (mean) and variable components:
$x(t) = \bar{x} [ 1 + \delta x(t)]$ and $o(t) = \bar{o} [1 + \delta o(t)]$, where $\delta x$ and $\delta o$ are the relative
deviations from the mean.
Two components contribute to the variations of the optical radiation: the synchrotron emission from the hot flow $\delta s$ 
and the radiation coming from reprocessing of the X-rays in the accretion disk $\delta d$
\begin{equation}
 \delta o(t) = \delta s(t) \frac{\bar{s}}{\bar{o}} + \delta d(t) \frac{\bar{d}}{\bar{o}}.
\end{equation}
The anti-correlation between synchrotron and the X-rays (see Section~\ref{sect:specmod}) can be represented as 
\begin{equation}\label{eq:s}
 \delta s(t) = - \frac{\rm rms_s}{\rm rms_x} \delta x(t),
\end{equation}
where rms denotes the fractional root mean square amplitude of variability of the corresponding components.\footnote{We assume
here that the decrease of the synchrotron luminosity is simultaneous with the increase of the X-rays,
neglecting possible delays (see the discussion in Section~\ref{sect:sumdis}).} 
The reprocessed radiation from the disk is a convolution of the X-ray light curve with the disk response function $r(t)$:
\begin{equation}\label{eq:d}
 \delta d(t) = \int\limits_{-\infty}^{t} r(t-t') \delta x(t') dt'.
\end{equation}
Assuming the entire disk luminosity in the optical band is due to irradiation and the response is a $\delta$-function, the
fractional rms of the disk would be equal to that of the X-rays.
Total optical light curve can be represented as
\begin{equation}\label{eq:o}
 \delta o(t) = \frac{\rm rms_s}{\rm rms_x} \frac{\bar{s}}{\bar{o}} 
    \left[- \delta x(t) + r_{\rm ds} \int\limits_{-\infty}^{t}\delta x(t') r(t-t')dt'\right],
\end{equation}
where we introduced the ratio of the absolute contributions of the  disk and the synchrotron to the variable optical component
(with the assumption of a $\delta$-function response): 
\begin{equation}
r_{\rm ds} = 
\frac{{\rm rms_x}} {{\rm rms_s}} \frac{\bar{d}}{\bar{s}} .
\end{equation}
As follows from Equation (\ref{eq:o}), the optical/X-ray CCF  contains two terms: one arising from synchrotron (it is just an
ACF of the X-rays with a negative sign) and another from the disk.

The Fourier transforms\footnote{The Fourier transforms corresponding to 
$\delta x(t)$, $\delta o(t)$, $\delta s(t)$, $\delta d(t)$, and $r(t)$ are
denoted as $X(f)$, $O(f)$, $S(f)$, $D(f)$, and $R(f)$, respectively, with $f$ being the Fourier frequency.}  
for the optical light curves are the sum of the synchrotron $S(f)\propto - X(f)$ and the disk 
$D(f) = X(f)R(f)$ transforms, and according to Equation~(\ref{eq:o}) can be represented as 
\begin{equation}
 O(f) = S(f) + D(f)\propto  X(f) \left[ -1 + r_{\rm ds} R(f)\right].
\end{equation}
The optical/X-ray cross-spectrum is proportional to the X-ray  power spectral density (PSD, $P_{\rm X}$) 
\begin{equation}
 C(f) = X^*(f) O(f) \propto  P_{\rm X}(f) \left[ -1 + r_{\rm ds} R(f) \right].
\end{equation}
Its phase is the  phase lag $\Delta \phi(f)$ and the time lag is $\Delta t(f) = \Delta \phi(f)/2\pi f$. 
The optical PSD consists of three terms: the synchrotron term, the disk term, and a cross term
\begin{equation}\label{eq:psdo}
 P_{\rm O}(f) \propto P_{\rm X}(f) \left\{1 + r_{\rm ds}^2\left| R(f)\right|^2 - 2r_{\rm ds}{\rm Re}\left[R(f)\right]\right\}.
\end{equation}

The model is fully determined by the rms in the X-ray and optical bands (${\rm rms_x}$ and ${\rm rms_o}$),
the ratio $r_{\rm ds}$, the disk response function $r(t)$ and the shape of the X-ray 
PSD.\footnote{We choose such a PSD normalization so that the integral over positive frequencies gives the square of relative rms
of the light curve \citep{MiKi89}  $\int P_{\rm X}(f) df = {\rm rms_x^2}$.}

\begin{figure*}
\plotone{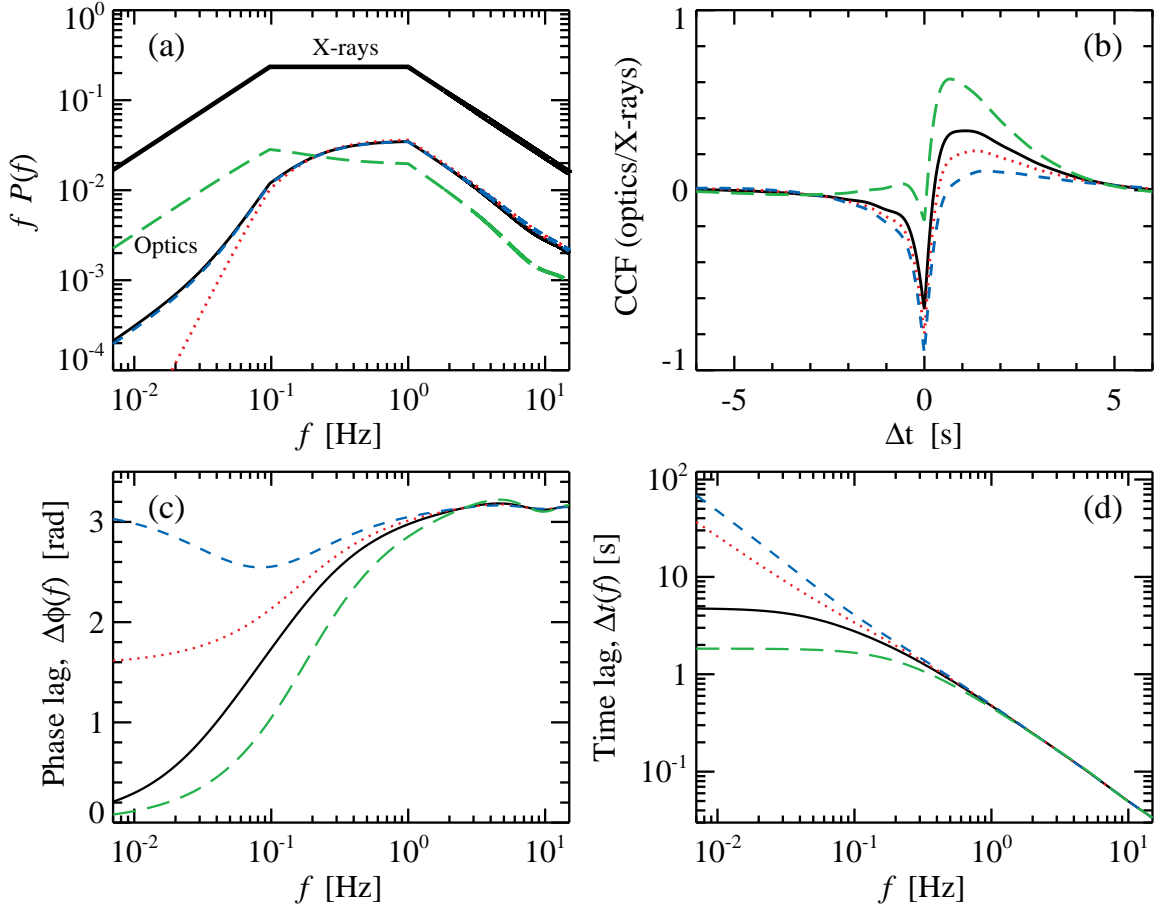}
\caption{
(a) X-ray and optical PSDs, (b) optical/X-ray CCFs, (c) optical/X-ray phase lags, and (d) time lags for various $r_{\rm ds}$. 
The results for $r_{\rm ds}=$0.7, 1.0, 1.3, and 2.5 are shown by dashed, dotted, solid and long-dashed lines, respectively. 
The X-ray PSD is shown with the thick solid line. Other parameters are as in  Figure~\ref{pic:result}. 
}\label{pic:params}
\end{figure*}

\subsection{Examples}
\label{sect:time_exam}

For an illustration, we choose the double-broken power-law X-ray PSD, 
$P_{\rm X}(f)\propto$ $f^{0}$, $f^{-1}$ and $f^{-2}$ with the breaks at 0.1 and 1.0 Hz. 
This shape is typical for BHBs in their hard state \citep{GCR99,RGC01,ABL05,GMD08}. 
We take ${\rm rms_x}=1.0$ and ${\rm rms_o}=0.3$ and consider a simple exponential response function 
\begin{equation}
  r(t) = \left\{
 \begin{array}{cc}
  \exp\left[ -(t-\tau_1)/\tau_2 \right]/\tau_2 , & t \geqslant \tau_1, \\
  0,                                             & t < \tau_1,
 \end{array} \right.
\end{equation}
where $\tau_1$ corresponds to the delay time of the disk response and $\tau_2$ is the response width.
The corresponding Fourier image is $R(f) = \exp({\rm i}x_1)/(1-{\rm i}x_2)$, where $x_k = 2 \pi f \tau_k$.
We take $\tau_1=0.1$ s and $\tau_2=1$ s. 
The results of simulations are shown in Figure~\ref{pic:result}. 

Contributions of different terms and the total PSDs are shown in Figure~\ref{pic:result}(a). 
The synchrotron term (the first term in Equation (\ref{eq:psdo})) has the same shape as the X-ray PSD, 
the disk (the second term in Equation (\ref{eq:psdo})) acts as a low-pass filter, and 
the cross term (the last  term in Equation (\ref{eq:psdo})) changes sign, 
giving positive contribution to the high-frequency part and negative contribution to the low-frequency part.
The combined contribution of the two (synchrotron and disk) components to the optical light curve
strongly suppresses the low-frequency power (as they vary in anti-phase) and increases 
the high-frequency power making a bump at frequencies $\sim$0.1--1 Hz  
in the optical power spectrum \citep[similar to that found in GX 339--4 by][]{GDD10}.
The excess of power at higher frequencies makes the optical ACF narrower than that of the X-rays.

The simulated CCFs for separate components and their joint contribution are shown in Figure~\ref{pic:result}(b).
The width of the precognition dip depends on the shape of the X-ray PSD.  
The amplitudes of both negative and positive peaks are reduced, when there are two components in the optical band, as they enter
Equation~(\ref{eq:o}) with different signs and thus partially cancel each other.
The fast rise of the CCF at zero lag is related to the interplay of the two optical components.  
The shape of the CCF resembles those found in Swift J1753.5--0127 \citep{DGS08,DSG11}.

The corresponding optical/X-ray phase and time lags are shown in Figure~\ref{pic:result}(c) and (d). 
For the assumed disk response function, the phase lag can be described by the analytical expression  
\citep[Equation 29 in][]{Pou02}:
\begin{equation}
 \tan \Delta \phi(f) = \frac {r_{\rm ds} (\sin x_1 + x_2 \cos x_1)}{-1 - x_2^2 + r_{\rm ds} (\cos x_1 - x_2 \sin x_1)}.
\end{equation}
If synchrotron dominates the optical emission  
(i.e.,  $r_{\rm ds} < 1$), then $\Delta \phi(f) \approx \pi$ and $\Delta t(f) \propto f^{-1}$ (dashed lines). 
Such a dependence was observed in XTE J1118+480 \citep{MBSK03}.  
In a general case, the synchrotron dominates at high frequencies, while at low frequencies we get 
\begin{equation} 
 \Delta \phi(f) \approx
\left\{ 
\begin{array}{cc}
\displaystyle \frac {r_{\rm ds}}{r_{\rm ds} - 1} (x_1+x_2) , &   \mbox{if}\ r_{\rm ds} > 1 ,   \\
\displaystyle \pi - \frac {r_{\rm ds}}{1 - r_{\rm ds}} (x_1+x_2) ,  &    \mbox{if}\ r_{\rm ds} <1   .
\end{array}
\right.
\end{equation}
In the first case ($r_{\rm ds} >1$), this translates into constant time lags 
$\Delta t \approx (\tau_1+\tau_2) r_{\rm ds}/(r_{\rm ds} - 1)$.

If the X-ray PSD is fixed, three parameters control the shape of the CCF and the optical PSD: $\tau_1$, $\tau_2$, and 
$r_{\rm ds}$. 
The first two parameters do not affect the CCF dramatically, therefore we further study the role of $r_{\rm ds}$. 
An increase of $r_{\rm ds}$ obviously results in suppression of the precognition dip in the CCF
(Figure~\ref{pic:params}(b)), in the disk-dominating regime $r_{\rm ds}\gg1$ giving the CCF of simple reprocessing.
At the same time, the role of this parameter in the PSD is not so straightforward, as can be seen in Figure~\ref{pic:params}(a).
The low-frequency tail is maximally suppressed at $r_{\rm ds}=1.0$, at which the disk and synchrotron nearly cancel each other. 
The high-frequency part is dominated by the synchrotron PSD and is almost independent of $r_{\rm ds}$. 
We note that for $r_{\rm ds}=1\pm \alpha$  (with $| \alpha |<1$)  the  PSD shapes are nearly the same, 
while the corresponding CCFs are different (compare the dashed and the solid lines in Figure~\ref{pic:params}(a) and (b)). 
The phase- or time-lag spectra (Figure~\ref{pic:params}(c) and (d)) also strongly depend on $r_{\rm ds}$.
The larger the value of $r_{\rm ds}$, the closer the phase spectrum is to that of simple reprocessing.

\section{Summary and discussion}\label{sect:sumdis}

In this Letter, we present a spectral model capable of explaining optical/X-ray timing features observed in LMXBs.
We argue that the optical emission is partially produced in the hot accretion flow (at distance $R\gtrsim30R_{\rm S}$)
by synchrotron radiation, which also provides seed photons for Comptonization.
Additional contribution to the optical band comes from reprocessing of the X-ray emission in the cold accretion disk.
We show that the presence of the two components can explain the observed shape of the CCF with the pronounced 
optical precognition dip. 
The model also reproduces another interesting feature observed in the LMXBs, namely, the optical ACF being narrower than the
X-ray one.
The explanation of the feature comes from the optical PSD, where the low-frequency part is suppressed and a bump appears 
at frequencies $\sim$0.1$-$1 Hz, both due to the presence of the cross term.
The  power at high frequencies originates from the interplay of the two optical components, 
which also results in the fast rise of the CCF at zero lag.
The Fourier frequency-dependent time lags resemble the observed ones.

We note that the proposed model is simplified, as it does not account for the fact that the optical synchrotron 
radiation comes from the outer parts of the hot flow, while the X-rays are likely to originate from the very vicinity of the
compact object.
This consideration leads to two complications: first, the synchrotron should lead the X-rays by the viscous timescale of 
about $\Delta t\sim 0.1$ s; second, it should have less power at high frequencies comparing to the X-rays. 
The latter is rather important as it leads to suppression of the high-frequency optical PSD, thus broadening the
precognition dip. 
The correct shape of the X-ray PSD is essential for comparison to the observed CCF, as it influences the shape of the
precognition dip dramatically, at the same time affecting the width of the optical peak at positive lags.
Generally, the broader is the X-ray ACF, the broader are the two features in the CCF
\citep[as indeed is observed,][]{DGS08,DSG11}.
The detailed comparison with the data will be a subject of our further investigations.

\acknowledgments
This work was supported by the Finnish Graduate School in Astronomy and Space Physics (AV), 
the Academy of Finland grant 127512 (JP), the Wihuri Foundation and ERC Advanced Research Grant 227634 (IV).  
The authors thank Mike Revnivtsev and Piergiorgio Casella for useful discussions. 


\end{document}